\documentclass[twocolumn,showpacs,prl,superscriptaddress,floatfix]{revtex4}   

\pdfoutput=1
\usepackage{graphicx,epsfig}

\newcommand{ \be }{\begin{eqnarray}}       
\newcommand{ \ee }{\end{eqnarray}}       
\newcommand{ \la }{\langle}       
\newcommand{ \ra }{\rangle}

\newcommand{ \mean }[1]{\left\langle #1 \right\rangle}   

\def\snn{$\sqrt{s_{NN}}$}

\def\P{$\cal P$}

\newcommand{ \psirp }{\Psi_{RP}}
\newcommand{ \phia }{\phi_{\alpha}}
\newcommand{ \phib }{\phi_{\beta}}
\newcommand{ \corr }{\mean{\cos(\phia+\phib-2\psirp)}}   
\usepackage{color}   
\definecolor{orange}{cmyk}{0.,0.353,1.,0.}    
\definecolor{dgreen}{cmyk}{1.,0.,1.,0.6}	

\usepackage{fixmath} 
\usepackage{dcolumn} 
\usepackage{bm} 

\begin{document}

\title{
Azimuthal Charged-Particle Correlations and Possible
Local Strong Parity Violation  
} 

\affiliation{Argonne National Laboratory, Argonne, Illinois 60439, USA}
\affiliation{University of Birmingham, Birmingham, United Kingdom}
\affiliation{Brookhaven National Laboratory, Upton, New York 11973, USA}
\affiliation{University of California, Berkeley, California 94720, USA}
\affiliation{University of California, Davis, California 95616, USA}
\affiliation{University of California, Los Angeles, California 90095, USA}
\affiliation{Universidade Estadual de Campinas, Sao Paulo, Brazil}
\affiliation{University of Illinois at Chicago, Chicago, Illinois 60607, USA}
\affiliation{Creighton University, Omaha, Nebraska 68178, USA}
\affiliation{Czech Technical University in Prague, FNSPE, Prague, 115 19, Czech Republic}
\affiliation{Nuclear Physics Institute AS CR, 250 68 \v{R}e\v{z}/Prague, Czech Republic}
\affiliation{Institute of Physics, Bhubaneswar 751005, India}
\affiliation{Indian Institute of Technology, Mumbai, India}
\affiliation{Indiana University, Bloomington, Indiana 47408, USA}
\affiliation{University of Jammu, Jammu 180001, India}
\affiliation{Joint Institute for Nuclear Research, Dubna, 141 980, Russia}
\affiliation{Kent State University, Kent, Ohio 44242, USA}
\affiliation{University of Kentucky, Lexington, Kentucky, 40506-0055, USA}
\affiliation{Institute of Modern Physics, Lanzhou, China}
\affiliation{Lawrence Berkeley National Laboratory, Berkeley, California 94720, USA}
\affiliation{Massachusetts Institute of Technology, Cambridge, MA 02139-4307, USA}
\affiliation{Max-Planck-Institut f\"ur Physik, Munich, Germany}
\affiliation{Michigan State University, East Lansing, Michigan 48824, USA}
\affiliation{Moscow Engineering Physics Institute, Moscow Russia}
\affiliation{City College of New York, New York City, New York 10031, USA}
\affiliation{NIKHEF and Utrecht University, Amsterdam, The Netherlands}
\affiliation{Ohio State University, Columbus, Ohio 43210, USA}
\affiliation{Old Dominion University, Norfolk, VA, 23529, USA}
\affiliation{Panjab University, Chandigarh 160014, India}
\affiliation{Pennsylvania State University, University Park, Pennsylvania 16802, USA}
\affiliation{Institute of High Energy Physics, Protvino, Russia}
\affiliation{Purdue University, West Lafayette, Indiana 47907, USA}
\affiliation{Pusan National University, Pusan, Republic of Korea}
\affiliation{University of Rajasthan, Jaipur 302004, India}
\affiliation{Rice University, Houston, Texas 77251, USA}
\affiliation{Universidade de Sao Paulo, Sao Paulo, Brazil}
\affiliation{University of Science \& Technology of China, Hefei 230026, China}
\affiliation{Shandong University, Jinan, Shandong 250100, China}
\affiliation{Shanghai Institute of Applied Physics, Shanghai 201800, China}
\affiliation{SUBATECH, Nantes, France}
\affiliation{Texas A\&M University, College Station, Texas 77843, USA}
\affiliation{University of Texas, Austin, Texas 78712, USA}
\affiliation{Tsinghua University, Beijing 100084, China}
\affiliation{United States Naval Academy, Annapolis, MD 21402, USA}
\affiliation{Valparaiso University, Valparaiso, Indiana 46383, USA}
\affiliation{Variable Energy Cyclotron Centre, Kolkata 700064, India}
\affiliation{Warsaw University of Technology, Warsaw, Poland}
\affiliation{University of Washington, Seattle, Washington 98195, USA}
\affiliation{Wayne State University, Detroit, Michigan 48201, USA}
\affiliation{Institute of Particle Physics, CCNU (HZNU), Wuhan 430079, China}
\affiliation{Yale University, New Haven, Connecticut 06520, USA}
\affiliation{University of Zagreb, Zagreb, HR-10002, Croatia}

\author{B.~I.~Abelev}\affiliation{University of Illinois at Chicago, Chicago, Illinois 60607, USA}
\author{M.~M.~Aggarwal}\affiliation{Panjab University, Chandigarh 160014, India}
\author{Z.~Ahammed}\affiliation{Variable Energy Cyclotron Centre, Kolkata 700064, India}
\author{A.~V.~Alakhverdyants}\affiliation{Joint Institute for Nuclear Research, Dubna, 141 980, Russia}
\author{B.~D.~Anderson}\affiliation{Kent State University, Kent, Ohio 44242, USA}
\author{D.~Arkhipkin}\affiliation{Brookhaven National Laboratory, Upton, New York 11973, USA}
\author{G.~S.~Averichev}\affiliation{Joint Institute for Nuclear Research, Dubna, 141 980, Russia}
\author{J.~Balewski}\affiliation{Massachusetts Institute of Technology, Cambridge, MA 02139-4307, USA}
\author{O.~Barannikova}\affiliation{University of Illinois at Chicago, Chicago, Illinois 60607, USA}
\author{L.~S.~Barnby}\affiliation{University of Birmingham, Birmingham, United Kingdom}
% ...

\collaboration{STAR Collaboration}\noaffiliation

\begin{abstract}
Parity-odd domains, corresponding to non-trivial topological solutions
of the QCD vacuum, might be created during relativistic heavy-ion
collisions.  These domains are predicted to lead to charge separation
of quarks along the system's orbital momentum axis.    
We investigate a three particle  
azimuthal correlator which is a
\P~even observable,  but
directly  sensitive  to  the  charge  separation  effect.
We  report  measurements 
of charged hadrons near center-of-mass rapidity 
with this observable 
in Au+Au and Cu+Cu collisions at $\sqrt{s_{NN}}$=200~GeV
using the  STAR detector.
A signal  consistent  with several expectations from
the theory is detected.    
We  discuss possible contributions  
from other effects  that are not  related to  parity  violation.  
\end{abstract}

\maketitle

Parity (\P) violation in the weak interaction was observed 
in 1957~\cite{Wu:1957my}.  However, until recently, parity 
has been thought to be
conserved in the strong interaction.  Modern QCD theory 
does allow for parity violation, but experiments have not seen 
this violation and the resulting constraints are 
tight~\cite{Peccei:2006as,Baker:2006ts}.
Recently, it has been suggested that the hot and dense matter created in
heavy-ion collisions may form metastable domains where the parity and
time-reversal symmetries are locally violated~\cite{Kharzeev:1998kz}.  In non-central
collisions, these domains may manifest themselves by giving positively
and negatively charged particles opposite-direction momentum 'kicks'
along the angular momentum vector of the collision.  The resulting
charge separation is a consequence
of two factors~\cite{Kharzeev:2004ey,Kharzeev:2007jp,Fukushima:2008xe}: the 
difference in numbers of quarks with 
positive and negative chiralities due to a non-zero topological charge of the
metastable region, and the interaction of these particles with the extremely
strong magnetic field produced in such a collision (leading to the effect
being called the ``Chiral Magnetic Effect'').
This separation of 
charges along the angular momentum vector 
would be a clear \P-violation.

The expectation from this local \P-violation is that the
relative sign of charge separation and angular momentum vectors
is random in each event.
This implies that any
\P-odd observable should yield zero when averaged over many events.  
An experimental search for this effect
must therefore involve comparing the measured charge separation signal
in each event with the expected fluctuations due to non-\P-violating
effects, or equivalently measuring correlations among particles in each 
event.
This Letter reports the result of such a search performed in 
200~GeV
Au+Au and Cu+Cu heavy-ion collisions 
with the STAR detector at the Relativistic Heavy Ion
Collider (RHIC).

\begin{figure}[tbp]
  \includegraphics[width=.46\textwidth]{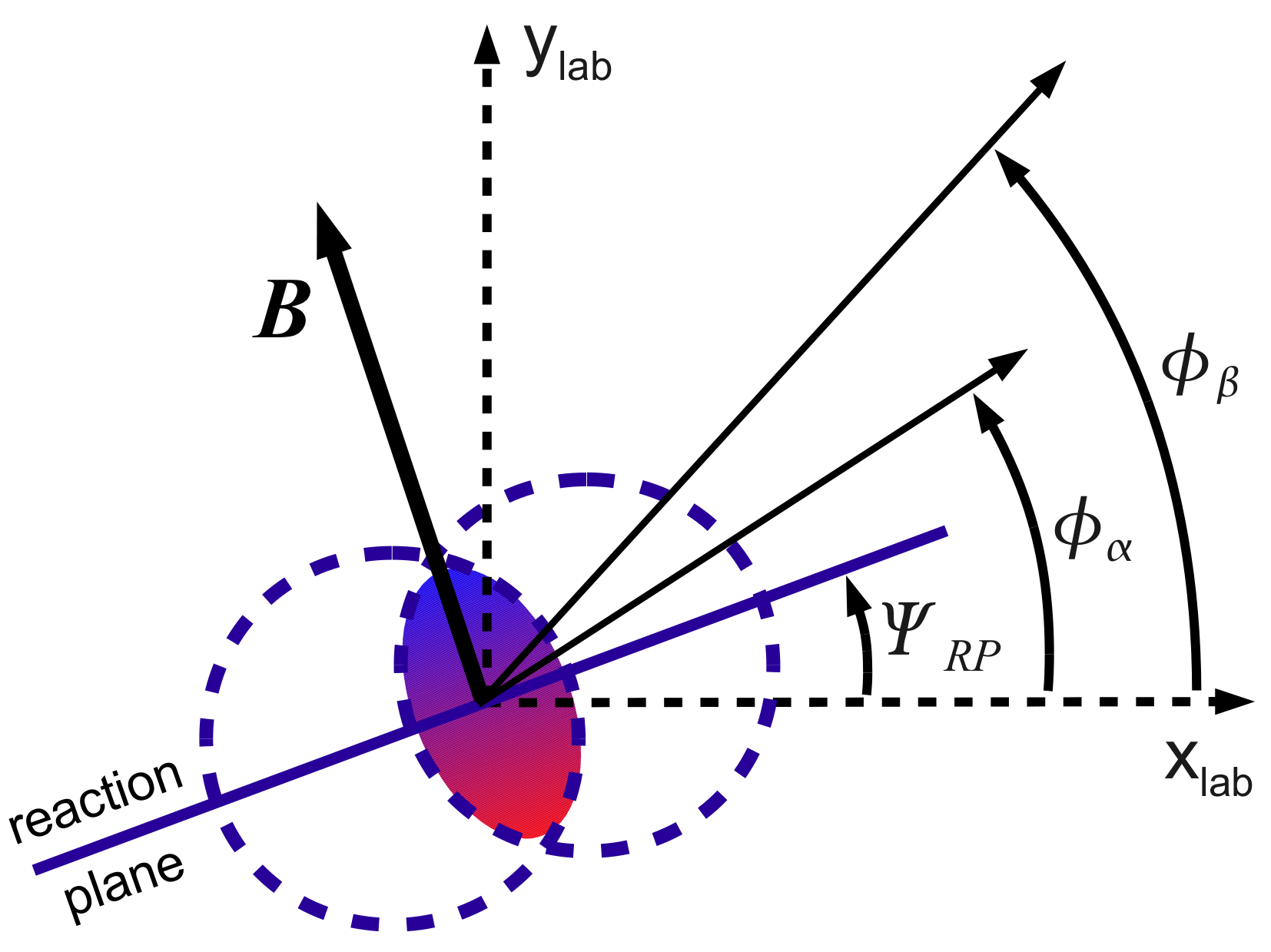}
  \caption{Schematic depiction of the transverse plane in a 
collision of two heavy ions (shown as dotted outlines - one emerging 
from and one going into the page).  The
azimuthal angles of the reaction plane and produced particles with charges 
$\alpha$ and $\beta$ as used in Eqs.~\ref{eq:expansion} and~\ref{e3p} 
are depicted here.
}
  \label{fig:cartoon}
\end{figure}

\emph{\textbf{Observables and Method.}}
In an event, charge separation along the angular momentum vector 
may be described phenomenologically
by sine terms in the Fourier decomposition of the charged particle
azimuthal distribution 
\be
 \frac{dN_\pm}{d\phi} &\propto& 1 + 2 v_1 \cos(\phi-\psirp)+
2 v_2 \cos(2(\phi-\psirp))+... 
\nonumber \\
&+&2 a_{\pm} \sin(\phi-\psirp) +...,
\label{eq:expansion}
\ee
where $\psirp$ is the azimuthal angle of the reaction plane
(the plane which contains the impact
parameter and beam momenta - see Fig.~\ref{fig:cartoon} for 
a schematic). $v_1$ and $v_2$ 
are coefficients accounting for the 
so-called directed and elliptic flow~\cite{Poskanzer:1998yz}. 
The $a$ parameters, $a_- = -a_+$, describe the \P-violating effect.
The spontaneous 
nature of the predicted parity violation means that the sign of 
$a_{+}$ and $a_{-}$ vary from event to event and on 
average $\mean{a_+}=\mean{a_-}=0$.

We may, however, expect non-zero values for the correlator 
$\mean{a_\alpha a_\beta}$ (where $\alpha$, $\beta$ represent 
electric
charge $+$ or $-$)
since \P-violating contributions to these 
observables would accumulate over many events.  
One could measure
$\mean{a_\alpha a_\beta}$ 
by calculating the average 
$\mean{\sin\Delta \phia\sin\Delta \phib}$ over
all particles of charge $\alpha$ paired with all 
particles of charge $\beta$
(here we have introduced the notation 
$\Delta \phi =\phi-\psirp$).
This is, however, also sensitive to 
several 
parity conserving 
physics backgrounds which produce 
correlations that have nonzero projections
along the angular momentum vector.   
This led to the proposal~\cite{Voloshin:2004vk} 
of the observable  
$\mean{ \cos(\phia +\phib -2\psirp) } = (
\mean{\cos\Delta \phia\cos\Delta \phib} 
-\mean{\sin\Delta \phia\sin\Delta \phib})$
which is sensitive to $-\mean{a_\alpha a_\beta}$. 
This observable
represents the difference between azimuthal correlations 
projected onto the direction
of the angular momentum vector and correlations projected onto the collision
reaction plane.  By taking this difference, 
these parity conserving 
correlations only produce backgrounds
inasmuch as they depend on 
orientation with respect to the reaction plane.  
As a consequence of the $\mean{\cos\Delta \phia\cos\Delta \phib}$
term, this observable is also sensitive to contributions from 
directed flow and its fluctuations.  Because the measurements presented 
in this Letter are for a 
symmetric rapidity region around center-of-mass rapidity, these 
contributions are negligible~\cite{Abelev:2008jga}.  

In practice, the reaction plane of a heavy-ion collision is not known, 
and one estimates it 
by measuring
the ``event plane'' 
which is reconstructed from particle azimuthal 
distributions~\cite{Poskanzer:1998yz}.
In the present analysis, 
this is done by using 3-particle azimuthal
correlations in which the third particle (labeled ``$c$'' below)
serves to measure the event plane~\cite{Voloshin:2004vk,PRCDRAFT}.
Assuming that the only correlation between 
particles of type ``$c$'' 
and particles $\alpha,\beta$ is through the common correlation 
to the reaction plane, we
can evaluate our observable 
for $-\mean{a_\alpha a_\beta}$ as
\be
\la \cos(\phi_a +\phi_\beta -2\psirp) \ra \, 
=
\la \cos(\phi_a +\phi_\beta -2\phi_c) \ra / v_{2,c}.
\label{e3p}
\ee  
By varying our choice of the type and momentum range of the
``$c$'' particles, we have tested this assumption as described below. 

Based on available theoretical understanding of
the Chiral Magnetic Effect 
we expect the following features of the correlator:
neglecting any ``final state'' interaction with
the medium, one expects 
$\mean{a_+a_+}=\mean{a_-a_-}=-\mean{a_+a_-} >0$. 
The likely effect of final state interactions in a hot
dense medium~\cite{Kharzeev:2007jp} is a 
suppression of back-to-back correlations 
(i.e. correlations among two particles that
travel in opposite directions through the medium), causing 
$\mean{a_+a_+} \gg |\mean{a_+a_-}|$. 
The dependence of the signal on the size of the colliding system has not 
yet been calculated, but 
one qualitative prediction is that the
suppression of opposite-charge correlations should be 
smaller in collisions of
lighter nuclei~\cite{Kharzeev:2007jp}. 
For a given collision system, under the assumption that the average 
size of the \P-violating domain does not change with 
centrality~\cite{Kharzeev:2007jp}, the correlator
should be inversely proportional to charged particle 
multiplicity, $N_{ch}$, 
scaled by a factor accounting
for the magnetic field in the collisions.
Finally, because the phenomenon is 
non-perturbative in nature,
we expect that the signal should not extend 
in transverse momentum far beyond $1$~GeV/c,
although this may
be affected by radial flow of the produced particles.

\emph{\textbf{Data and Detector.}}
The data were obtained with the STAR detector~\cite{Ackermann:2002ad} 
during RHIC runs in
2003-04 and 2004-05. 
The results are based on 14.7M Au+Au and 13.9M Cu+Cu
events at \snn=200~GeV.
A minimum bias trigger was used with events sorted into centrality classes
based upon charged particle multiplicity. 

The correlations are reported for charged particle tracks measured 
in the STAR Time Projection Chamber (TPC)
with pseudorapidity 
$|\eta|<1.0$ and transverse momentum $0.15<p_t<2$~GeV/c.  
For event plane determination,
in addition to the main TPC we use measurements in the two Forward TPCs 
($2.7<|\eta|<3.9$) and two Zero 
Degree Calorimeter Shower Maximum 
Detectors (ZDC-SMDs)~\cite{Adams:2005ca}.  The 
latter are sensitive to the directed flow of  
neutrons in the beam rapidity region.  

The STAR detector is well suited 
to measure azimuthal correlations.
The TPC has full azimuthal coverage and a 
charged particle track reconstruction
efficiency of approximately
85\%.  Nevertheless, TPC sector 
boundaries, 
occasional readout channel outages, 
etc., may
introduce biases in the analysis. In particular, they may 
cause inefficiencies that
are different for positive and negative particles. 
In evaluating Eq.~\ref{e3p}, we correct for
detector effects
(following~\cite{Selyuzhenkov:2007zi,Borghini:2002vp}) by
replacing $\cos (n\phi_i)$ with $[\cos (n\phi_i)$ - $\mean{ \cos (n\phi)}]$
for each particle, and similarly for the $\sin (n\phi_i)$
terms 
which also appear~\cite{PRCDRAFT}.
We calculate the ``re-centering'' corrections $\mean{\cos (n\phi)}$
and $\mean{\sin (n\phi)}$
as a function of time as well as event multiplicity and z-vertex position.  
We also account for the acceptance dependence on particle 
$\eta$, $p_t$ and charge. 
Higher order
acceptance corrections are 
found to be negligible.

\emph{\textbf{Experimental Uncertainties.}}
The dominant experimental systematic error comes from our knowledge of $v_2$ 
which is used in Eq. ~\ref{e3p}.  
The shaded bands in the figures
reflect this systematic error
with the actual points determined by applying $v_2$ for TPC particles
as measured using the reaction plane found in the FTPC.  

Other experimental systematics, including possible biases due to acceptance
and detector efficiency and errors related to track quality cuts, are found to
be comparable to or smaller than statistical errors~\cite{PRCDRAFT}. 

\begin{figure}[tbp]
  \includegraphics[width=.46\textwidth]{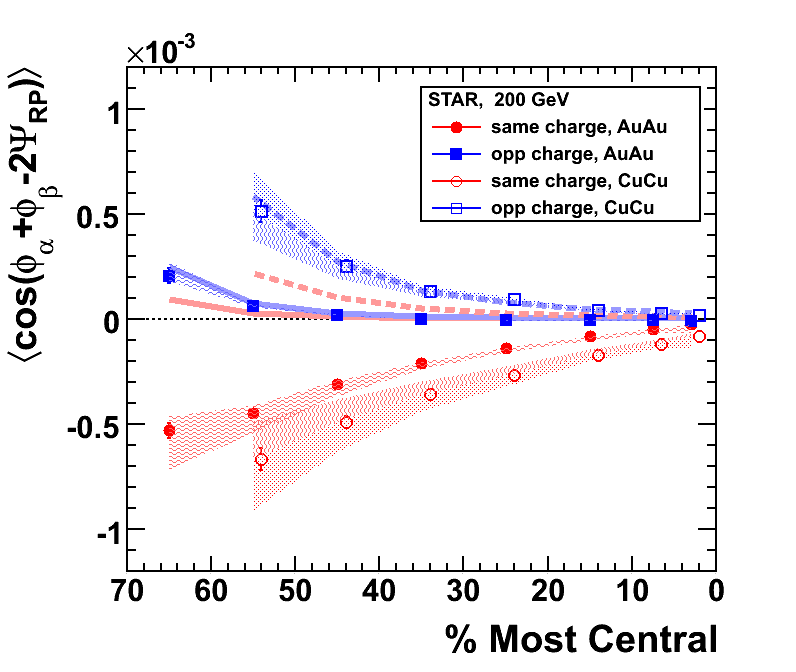}
  \caption{ $\mean{\cos(\phi_a +\phi_\beta -2\psirp) }$
in Au+Au and Cu+Cu
collisions at $\sqrt{s_{NN}}=200$~GeV  
calculated using Eq.~\ref{e3p}.
The thick solid (Au+Au) and 
dashed (Cu+Cu) lines represent HIJING calculations 
of the contributions from 3-particle correlations.   
Shaded bands represent uncertainty from the measurement of $v_2$. 
Collision centrality increases from left to right.
}
  \label{fig:uuv2_200}
\end{figure}

\emph{\textbf{Results.}}
Figure~\ref{fig:uuv2_200}
presents $\corr$ for Au+Au and Cu+Cu
collisions at \snn=200~GeV as evaluated using the
right-hand side of Eq.~\ref{e3p}
(error bars indicate statistical errors). 
The signal in Cu+Cu collisions is 
larger than the signal in Au+Au collisions at
the same centrality, qualitatively consistent 
with the expected decrease of the signal
with increasing multiplicity.
For the Au+Au system, opposite-charge correlations are clearly smaller in 
magnitude than same-charge correlations, in qualitative agreement with the possible
suppression of back-to-back charge correlations.
This is supported by the observation of a smaller 
difference in magnitude between
same-charge and opposite-charge correlations in the smaller Cu+Cu system.
However, there is a large potential background
contribution from 3-particle clusters to opposite-charge correlations which 
is discussed below
and indicated 
by the thick solid (Au+Au) and dashed (Cu+Cu) lines on Fig.~\ref{fig:uuv2_200}. 


\begin{figure}[tbp]
  \includegraphics[width=.46\textwidth]{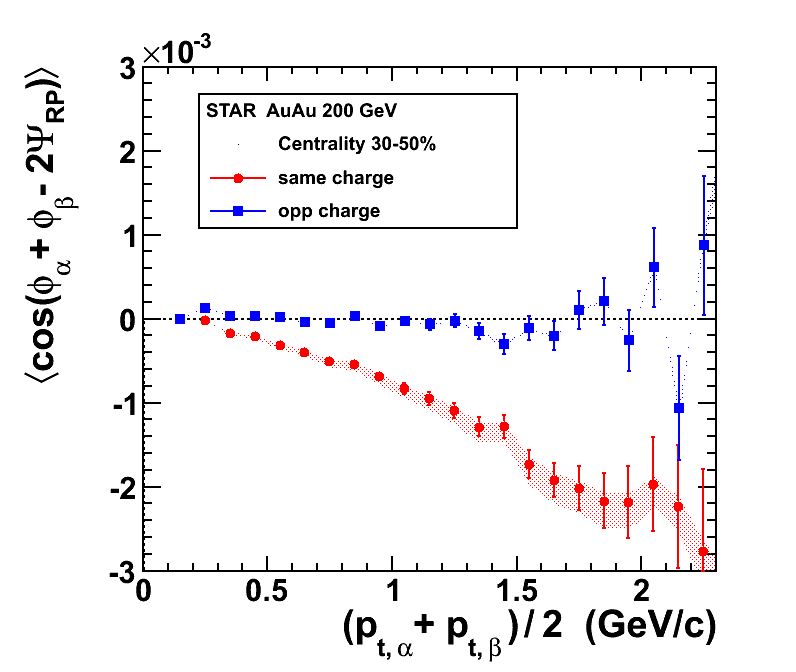}
  \caption{Dependence of $\corr$ on $\frac{1}{2}(p_{t,\alpha}+p_{t,\beta})$ 
calculated using no upper cut on particles' $p_t$.  Shaded bands represent $v_2$
uncertainty.
} 
  \label{fig:uuv2diffb}
\end{figure}

Figure~\ref{fig:uuv2diffb} shows the 
dependence of the signal on the
sum of the transverse momentum magnitudes of the 
two particles for
the 30-50\% centrality range in 200~GeV Au+Au collisions.
We do not observe a 
saturation or drop of the magnitude of the signal at high $p_t$ as 
one might naively expect
for local \P~ violation.  
The correlations are nearly 
independent of the $p_t$ difference
over the range 
$0 < |p_{t,\alpha}-p_{t,\beta}| < 2$~GeV/c~\cite{PRCDRAFT}, which
excludes quantum interference (HBT) or
Coulomb effects as possible explanations for the signal.
We have studied the dependence of the signal on
$|\eta_\alpha -\eta_\beta|$ \cite{PRCDRAFT}, and find
that the signal has a width of about one unit of
$\eta$.

\begin{figure}[tbp]
 \includegraphics[width=.48\textwidth]{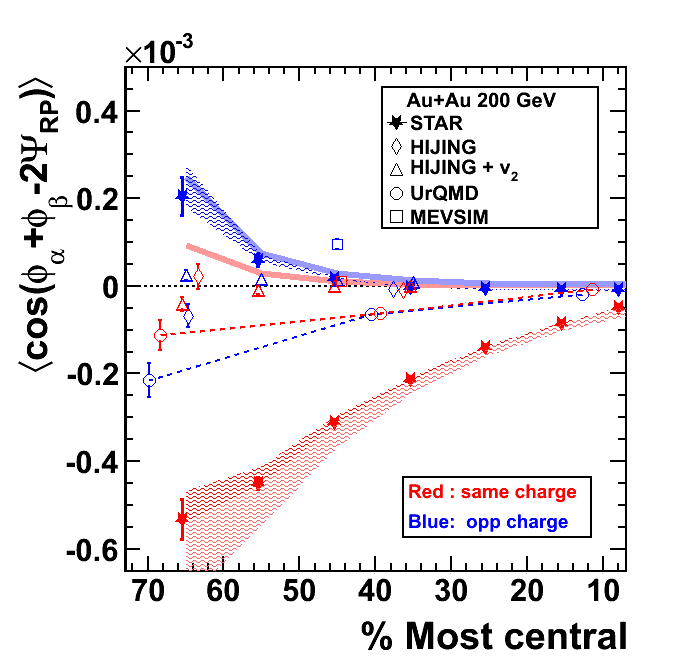}
  \caption{$\corr$ results from 200~GeV Au+Au collisions are compared
to calculations with 
event generators HIJING (with and without an ``elliptic flow
afterburner''),UrQMD
(connected by dashed lines), 
and MEVSIM.
Thick lines represent HIJING reaction-plane-independent
background.
}
  \label{fig:AuAusimulations}
\end{figure}

\emph{\textbf{Physics Backgrounds.}}
We first consider backgrounds due to multi-particle 
correlations (3 or more particles) 
which are not related to the reaction plane. 
This contribution affects the 
assumption that
two particle correlations with respect to the reaction plane
(L.H.S. of Eq.~\ref{e3p}) can be evaluated in practice via
three particle correlations (R.H.S. of  Eq.~\ref{e3p}).
Evidence supporting this 
assumption 
comes from the consistency of 
same-charge results when the reaction plane is found
using particles ``$c$'' detected in the TPC, FTPC, or ZDC-SMD,
though the FTPC and (particularly) ZDC-SMD analyses have large
statistical errors in the most peripheral bins.  
This multi-particle background should be negligible 
when the ZDC-SMD event plane is used, 
so it can certainly be reduced and this 
is an
important goal of 
future high statistics runs.
To study these backgrounds in the current analysis, we use the heavy-ion 
event model HIJING~\cite{refHIJING} 
(used with default settings and jet quenching 
off in all calculations shown in this Letter)
which includes production and fragmentation of mini-jets.  
We find that the contribution to opposite-charge correlations of
three particle correlations in HIJING (represented by the thick solid and dashed
lines in Fig.~\ref{fig:uuv2_200} and~\ref{fig:AuAusimulations}) 
is similar to the measured signal in several peripheral bins.  
We thus cannot conclude that
there is an opposite-charge signal above possible background.  The same-charge 
signal predicted by three-particle correlations in HIJING
is much smaller and of opposite sign compared to that
seen in the data.  

Another class of backgrounds (which cannot be reduced by better determination
of the reaction plane) consists of processes in which particles $\alpha$ and $\beta$ are
products of a cluster of two or more particles (for example a resonance decay or jet) 
and the cluster itself exhibits
elliptic flow
or fragments differently when emitted in-plane compared 
to out-of-plane~\cite{Voloshin:2004vk,PRCDRAFT}.

For jets with a leading charged particle of $p_{t}>3$~GeV/c, we 
estimate the contribution using previous STAR
measurements~\cite{Adams:2003kv,Adams:2005ph} and find it to be negligible.
To extend the study to lower $p_t$, we rely on HIJING calculations 
of two particle correlations with respect to the true reaction plane.
These calculations also predict the contribution to be small compared to our 
measured signal as shown by the triangles in Fig.~\ref{fig:AuAusimulations}. 

Resonance decays have the potential to contribute to $\corr$.
In addition, previous correlation
measurements from the ISR~\cite{Foa:1975eu} 
and RHIC~\cite{Alver:2007wy,Daugherity:2008su}
indicate that a prominent
role in particle production is played by clusters.  A much smaller
signal is expected for same- than opposite-charge correlations from
resonances, which is qualitatively very unlike the signal 
shown in Fig.~\ref{fig:uuv2_200}.   Kinematic studies demonstrate 
that it is very difficult for the
correct sign of fake signal to be created in the same-charge
correlations without postulating a negative value of $v_2$ for the
resonances or 
particles from cluster decays.

To search for
other backgrounds to $\corr$,
we have simulated Au+Au collisions with heavy-ion event generators 
MEVSIM~\cite{refMEVSIM}, UrQMD~\cite{refUrQMD}, and HIJING 
(with and without an elliptic flow afterburner implemented as 
suggested in~\cite{Poskanzer:1998yz}) for comparison 
and these results (calculated using the true reaction plane in all cases) 
are shown as open symbols in Fig.~\ref{fig:AuAusimulations}.
MEVSIM only includes correlations due to resonance decays
and an overall elliptic flow pattern.  UrQMD and HIJING are real physics models 
of the collision and so include correlations
from many different physical processes.  
Figure~\ref{fig:AuAusimulations} shows that no generator gives 
qualitative agreement with data for two particle correlations with respect to 
the reaction plane.  
The models also do not match the 
measured values for reaction plane independent 
correlations, ${\mean{\cos(\phia-\phib)}}$~\cite{PRCDRAFT}.

Other effects we find to produce insignificant contributions 
\cite{PRCDRAFT} include 
global polarization of 
hyperons along the direction of the system angular momentum. 


\emph{\textbf{Summary.}}
Measurements of three particle correlations that are directly
sensitive to predicted local \P-violation in heavy-ion collisions have
been presented for Au+Au and Cu+Cu collisions at $\sqrt{s_{NN}}$=200~GeV
as a function of collision centrality and (in Au+Au) 
particle transverse momentum,
with a more complete set of measurements reported in \cite{PRCDRAFT}.

Qualitatively the results agree 
with the magnitude and gross features
of the theoretical predictions
for local \P-violation in heavy-ion
collisions, but the signal persists to higher transverse
momentum than expected.
The observable used in our analysis is
\P-even and might be sensitive to non-parity-violating effects. 
So far, with the systematics checks discussed in this 
paper and \cite{PRCDRAFT}, 
we have not identified effects that would explain the
observed same-charge correlations.  
The observed signal cannot be described by the background models
that we have studied
(HIJING, HIJING+$v_2$, UrQMD, MEVSIM), 
which span a broad range of hadronic physics.

A number of future experiments and analyses are naturally suggested by
these results. One of them is the dependence of the
signal on the collision energy. 
The charge separation effect is expected to depend strongly on the formation 
of a quark-gluon plasma~\cite{Kharzeev:2007jp}, 
and the signal might be greatly suppressed or completely absent
at an energy below that at which a QGP can be formed.

Improved theoretical calculations of the expected signal and potential
physics backgrounds in high energy heavy-ion collisions are essential to
understand whether or not the observed signal is due to local strong
parity violation, and to further experimental study of this phenomenon.

\emph{\textbf{Acknowledgments.}}
We thank D.~Kharzeev for many helpful discussions.
We thank the RHIC Operations Group and RCF at BNL, the NERSC Center at 
LBNL and the Open Science Grid consortium for providing resources and 
support. This work was supported in part by the Offices of NP and HEP 
within the U.S. DOE Office of Science, the U.S. NSF, the Sloan 
Foundation, the DFG cluster of excellence `Origin and Structure of the 
Universe', CNRS/IN2P3, STFC and EPSRC of the United Kingdom, FAPESP CNPq 
of Brazil, Ministry of Ed. and Sci. of the Russian Federation, 
NNSFC, CAS, MoST, and MoE of China, GA and MSMT of the Czech Republic, 
FOM and NOW of the Netherlands, DAE, DST, and CSIR of India, Polish 
Ministry of Sci. and Higher Ed., Korea Research Foundation, Ministry of 
Sci., Ed. and Sports of the Rep. Of Croatia, Russian Ministry of Sci. 
and Tech, and RosAtom of Russia.



\medskip
\noindent
\begin{thebibliography}{99}

\bibitem{Wu:1957my}
  C.~S.~Wu, E.~Ambler, R.~W.~Hayward, D.~D.~Hoppes and R.~P.~Hudson,
  Phys.\ Rev.\  {\bf 105}, 1413 (1957).

\bibitem{Peccei:2006as}
  R.~D.~Peccei,
  Lect.\ Notes Phys.\  {\bf 741}, 3 (2008).

\bibitem{Baker:2006ts}
  C.~A.~Baker {\it et al.},
  Phys.\ Rev.\ Lett.\  {\bf 97}, 131801 (2006).

\bibitem{Kharzeev:1998kz}
  D.~Kharzeev, R.~D.~Pisarski and M.~H.~G.~Tytgat,
  Phys.\ Rev.\ Lett.\  {\bf 81}, 512 (1998).

\bibitem{Kharzeev:2004ey}
  D.~Kharzeev,
  Phys.\ Lett.\  B {\bf 633}, 260 (2006).

\bibitem{Kharzeev:2007jp}
  D.~E.~Kharzeev, L.~D.~McLerran and H.~J.~Warringa,
  Nucl.\ Phys.\  A {\bf 803}, 227 (2008).

\bibitem{Fukushima:2008xe}
  K.~Fukushima, D.~E.~Kharzeev and H.~J.~Warringa,
  Phys.\ Rev.\  D {\bf 78}, 074033 (2008).

\bibitem{Poskanzer:1998yz}
  A.~M.~Poskanzer and S.~A.~Voloshin,
  Phys.\ Rev.\  C {\bf 58}, 1671 (1998).

\bibitem{Voloshin:2004vk}
  S.~A.~Voloshin,
  Phys.\ Rev.\  C {\bf 70}, 057901 (2004).

\bibitem{Abelev:2008jga}
  B.~I.~Abelev {\it et al.},
  Phys.\ Rev.\ Lett.\  {\bf 101}, 252301 (2008).

\bibitem{PRCDRAFT}
  B.~I.~Abelev {\it et al.},
arXiv:0909.1717 [nucl-ex].

\bibitem{Ackermann:2002ad}
  K.~H.~Ackermann {\it et al.},
  Nucl.\ Instrum.\ Meth.\  A {\bf 499}, 624 (2003).




\bibitem{Adams:2005ca}
  J.~Adams {\it et al.},
  Phys.\ Rev.\  C {\bf 73}, 034903 (2006).

%

\bibitem{Selyuzhenkov:2007zi}
 I.~Selyuzhenkov and S.~Voloshin,
 Phys.\ Rev.\  C {\bf 77}, 034904 (2008).

\bibitem{Borghini:2002vp}
 N.~Borghini, P.~M.~Dinh and J.~Y.~Ollitrault,
 Phys.\ Rev.\  C {\bf 66}, 014905 (2002).

\bibitem{refHIJING} 
 M. Gyulassy and X.-N. Wang, Comput. Phys. Commun. {\bf 83},
  307 (1994);
  X.N. Wang and M. Gyulassy, Phys. Rev. D {\bf 44}, 3501 (1991).



\bibitem{Adams:2003kv}
  J.~Adams {\it et al.},
  Phys.\ Rev.\ Lett.\  {\bf 91}, 172302 (2003).

\bibitem{Adams:2005ph}
  J.~Adams {\it et al.},
  Phys.\ Rev.\ Lett.\  {\bf 95}, 152301 (2005).

\bibitem{Foa:1975eu}
  L.~Foa,
  Phys.\ Rept.\  {\bf 22} 1 (1975).

\bibitem{Alver:2007wy}
  B.~Alver {\it et al.},
  Phys.\ Rev.\  C {\bf 75}, 054913 (2007).

  \bibitem{Daugherity:2008su}
    M.~Daugherity,
    J.\ Phys.\ G {\bf 35}, 104090 (2008).

\bibitem{refMEVSIM}
  R.~L.~Ray and R.~S.~Longacre,
  arXiv:nucl-ex/0008009.

\bibitem{refUrQMD}
 S.~A.~Bass {\it et al.},
 Prog.\ Part.\ Nucl.\ Phys.\  {\bf 41}, 255 (1998).





\end{thebibliography}
\end{document}